# Interface Phonon Modes in the [AlN/GaN]$_{20}$ and [Al$_{0.35}$Ga$_{0.65}$N/Al$_{0.55}$Ga$_{0.45}$N]$_{20}$ 2D Multi Quantum Well Structures

A. K. Sivadasan,[*,1] Chirantan Singha,[2] A. Bhattacharyya,[3] and Sandip Dhara[*,1]

[1] Nanomaterials and Sensors Section, Surface and Nanoscience Division, Indira Gandhi Centre for Atomic Research, Homi Bhabha National Institute, Kalpakkam–603102, India

[2] Centre for Research in Nanoscience and Nanotechnology, University of Calcutta, JD2, Sector–III, Saltlake City, Kolkata–700106, West Bengal, India

[3] Institute of Radio Physics and Electronics, University of Calcutta, 92, A.P.C. Road, Kolkata–700009, West Bengal, India

E-mail: sivankondazhy@gmail.com ; dhara@igcar.gov.in

Interface phonon (IF) modes of $c$–plane oriented [AlN/GaN]$_{20}$ and [Al$_{0.35}$Ga$_{0.65}$N/Al$_{0.55}$Ga$_{0.45}$N]$_{20}$ multi quantum well (MQW) structures grown via plasma assisted molecular beam epitaxy are reported. The effect of variation in dielectric constant of "barrier" layers to the IF optical phonon modes of "well" layers periodically arranged in the MQWs are investigated.

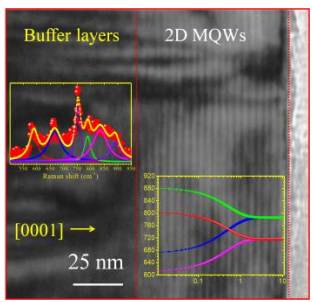

The observation of interface (IF) phonon modes in the recorded Raman spectra of $c$–plane oriented [AlN/GaN]$_{20}$ and [Al$_{0.35}$Ga$_{0.65}$N/Al$_{0.55}$Ga$_{0.45}$N]$_{20}$ multi quantum well (MQW) structures grown via plasma assisted molecular beam epitaxy are reported. The nominal shift in IF phonon mode of $E_1$ symmetry for [Al$_{0.35}$Ga$_{0.65}$N/Al$_{0.55}$Ga$_{0.45}$N]$_{20}$ compared to that of the [AlN/GaN]$_{20}$ MQW structure is understood on the basis of change in dielectric constants ($\varepsilon_m$) of the surrounding medium. The presence of buffer layers in [Al$_{0.35}$Ga$_{0.65}$N/Al$_{0.55}$Ga$_{0.45}$N]$_{20}$ MQW over the sapphire substrate is also understood by characterizing the IF phonon mode of $A_1$ symmetry. The observed IF phonon modes in the spectra are attributed to the relaxation of Raman selection rules away from the Brillouin zone centre because of the breakdown of translational symmetry of surface potential due to the presence of the periodic interfaces and surface modulations in the superlattice structures of MQWs. The corresponding required edge lengths ($L$) of 2D plates, for the observation of the breakdown of surface potential, are computed from the simulated dispersion relation curve of IF modes. The integral multiples of uniformly distributed platelets, originated due to the horizontal uneven irregularities on the surface of superlattices, are matched with the calculated $L$ values.

## 1. Introduction

An analogue to the crystallography, the superlattice (SL) is termed as a "lattice of lattices". Thus, any regular pattern or sequence of semiconducting layered heterostructures can be considered as a SL. A quantum well (QW) is a heterostructure of two or more semiconducting material in which one thin "well" layer is sandwiched by two other "barrier" layers. Consequently, the difference in electronic configurations in the semiconducting materials generates a finite "potential well" in which charge carriers of the material posses a lower energy with discrete levels instead of continuum as compared to that of with bulk materials. If the SL possesses a significant wave function penetration between the adjacent wells due to the tunneling effects, then the structure is called as a multi quantum well (MQW).[1-3] Similarly, an analogous to the "bands" in the crystal structure; if the QWs are arranged very close to each other with a thickness of the order of a few nanometers, then the significant wave function penetration between the QWs may lead to the creation of "mini-bands" in the SL structures. The electric field perpendicular to the MQW layers leads to the splitting of energy levels and the corresponding phenomena is called as the quantum-confined Stark effect (QCSE). The major applications of MQWs in the optoelectronic devices are waveguide modulator, self electronic effect devices (SEED), optical switching, MQW light emitting diodes (LEDs) and laser diodes (LDs).[1-3] Among them, the group III nitride multi quantum wells (MQWs) based nanostructures have remarkable applications in LEDs and LDs, which fall in the visible to ultraviolet (UV) region of the electro-magnetic spectra.[1-5] Because of the capability to create two dimensional (2D) electron gas at the heterojunctions, the III–nitrides especially AlGaN/GaN MQWs and SLs are used for developing high electron mobility transistors (HEMT) along with the heterojunction field effect transistors (HJFET) and bipolar transistors.[4-9] The SLs and MQWs of AlGaN/GaN with different Al percentages are used as active region, as well as buffer, carrier confining and electron blocking layers (EBL) to improve the internal and external quantum efficiency of blue, UV and deep–UV LEDs and LDs.[4-9]

The Raman spectroscopy demands the wave vector $q \approx 0$ selection rule for the observation of phonon modes in the infinite crystals as a consequence of the infinite periodicity of the crystal lattice. Whereas, in the case of optical phonons in polar semiconductor SLs and MQWs; the approximation of negligible crystal-momentum transfer for the backscattered waves as in the case of bulk crystal may not be true for certain phonon modes of SLs or MQWs with a larger periodicity. Moreover, the semiconductor SLs, single-quantum well structures, and MQWs are famous 2D systems because of the confinement of phonons and charge carriers within a plane where the degree of phonon confinement is 1D, as the confinement is restricted along one direction which is perpendicular to the SL plane.[1-3] However, the behavior of the different optical phonon modes at the interface (IF) of MQW structures and SLs are very much exciting and speculating topic for the current scientific research area of group III nitride polar wurtzite structures. Owing to the boundary conditions for the optical modes in the MQWs and the nature of parallel and perpendicular component of frequency dependent dielectric constants in the quantum confined layers, the solutions of Maxwell equations can take different mathematical functional forms leading to four types of optical phonon modes, namely, confined modes, IF modes, propagating modes and half–space modes.[9,10] The electron confinement in the MQW structures also can influence the structural and optical properties of the materials.[11] The allowed Raman modes of wurtzite group III nitrides are well studied, even though, there are few reports on specific class of phonon modes which can be localized on the surfaces of nanostructures. The frequency of such lattice vibrations are observed in between the transverse and longitudinal optical ($(TO)_{q=0}$ and $(LO)_{q=0}$) phonon modes named as IF or surface optical (SO) phonon modes.[12-14] The effect of surface–modulation–assisted electron–SO phonon scattering is suggested to be responsible for the prominent appearance of SO phonon modes and hence it is also known as Fröhlich modes.[11-22] The SO, as well as IF phonon modes are the vibrations of surface atoms whose amplitudes are confined near to the surface region of the materials. The SO or IF phonon modes are observed because of the breakdown of translational symmetry of the surface potential in different periodic structures.[23-25] The breakdown of translational symmetry leads to an additional crystal momentum that allows the evolution of phonon mode away from the Brillouin zone center. The additional momenta generated for creating these non-zone centre phonons can be absorbed by the periodic structure of the lattice in terms of quantized units of $2\pi/\lambda$, where $\lambda$ is the periodicity of the SLs or MQWs. This phenomenon is not possible to observe in the bulk materials because of the momentum conservation due to their perfect crystal symmetry.[23-25] Therefore, the nanomaterials such as 1D nanowires, 2D MQWs and SLs with a large surface area contribute significantly towards the SO or IF phonon signal as compared to that for the bulk crystals.[26] It was well established that the SO phonon mode frequency depends on the dielectric constant of the surrounding medium ($\varepsilon_m$), as well as the shape, interfaces, saw tooth edges, periodicity of surface defects, diameter modulation and variation of density of the nanostructures.[12-28]

In the present study, we investigated the behavior of IF phonon modes in the Raman spectra of $[AlN/GaN]_{20}$ and $[Al_{0.35}Ga_{0.65}N/Al_{0.55}Ga_{0.45}N]_{20}$ MQW structures originating from the breakdown of translational symmetry of the surface potential because of the presence of multiple periodic structures and interfaces of MQWs. The wave vector ($q$) was calculated in accordance with the wavelength ($\lambda$) originated from the periodicity of MQWs for the IF phonons and the corresponding required edge lengths ($L$) of 2D plates were investigated using a theoretical simulation of the dispersion relation in the 2D MQW structures. We also studied the change in IF phonon frequency in the observed spectra with respect to the difference in dielectric constant of the surrounding medium for MQWs and bi–layer thin film.

## 2. Experimental Details

### 2.1 Synthesis of MQW structures

The growth of MQW structures of [GaN/AlN]$_{20}$ and [Al$_{0.35}$Ga$_{0.65}$N/Al$_{0.55}$Ga$_{0.45}$N]$_{20}$ SLs on the *c*–plane oriented sapphire substrate were carried out with the help of plasma assisted molecular beam epitaxy (PAMBE). A VEECO Gen 930 system with standard effusion cells for group III sources and an RF plasma source for N activation was employed. Growth was carried out on sapphire substrates using a three step process, including a high temperature nitridation step and deposition of a high–temperature grown AlN buffer layer before the growth of the thin films and quantum structures at 750 $^{o}$C. Group III–rich conditions were employed to ensure smooth interfaces between well and barrier. A large number of threading dislocations were reported in the as-prepared nucleation layers of SLs.[29,30] However, a reduction of the dislocation density was observed in the sample during the deposition of the bulk AlGaN or AlN film (~300 nm) before starting the MQW deposition which might help to construct the active region relatively free from the dislocations. The detailed growth process and its basic characterizations were reported in our previous articles for studying the optoelectronic properties.[29,30]

### 2.1 Probing the vibrational properties of MQW structures

The vibrational properties of MQWs and bi–layer thin film were studied using Raman spectroscopy (inVia, Renishaw,UK). The Ar$^{+}$ laser with a wavelength of 514.5 nm was used as an excitation source. A grating with 1800 gr.mm$^{-1}$ ruling was used as a monochromatizer for the backscattered waves from the sample. The thermoelectrically cooled CCD detector was used in the backscattering mode to record the Raman shift corresponding to the Stokes lines. The sample excitations as well as the scattered spectra were collected using a 50X objective (Olympus 50X) with a numerical aperture (N.A.) value of 0.35 and a working distance of 18 mm. The complete schematic outline of the experimental setup is available in one of our report.[31] The corresponding laser beam diameter ($d$ = 1.22$\lambda$/N.A.) emanated from the objective for an excitation wavelength ($\lambda$) of 514.5 nm was ~1.80 µm. The Raman modes were recorded at low temperature from the sample kept in the liquid N cooled automated temperature controlled chamber (Linkam THMS 600) and the observed Raman modes were de–convoluted and peak fitted with Lorentzian function for further analysis. Transmission Electron Microscopy (TEM; Tecnai 20 G20) operating at 200 keV was used for typical cross–sectional microstructural studies of [Al$_{0.35}$Ga$_{0.65}$N/Al$_{0.55}$Ga$_{0.45}$N]$_{20}$ MQW.

## 3. Results and Discussions

### 3.1 Vibrational properties of MQW structures

The vibrational properties of the samples with MQW structures of *c*–plane oriented [GaN(~1.25 nm)/AlN(~2.0 nm)]$_{20}$ (schematic in Fig. 1(a)) and [Al$_{0.35}$Ga$_{0.65}$N(~1.75 nm)/Al$_{0.55}$Ga$_{0.45}$N(~2.0 nm)]$_{20}$ (Fig. 1(b) showing schematic as well as a TEM micrograph at the outset) SLs were probed by Raman spectroscopy at 80K. The periodic bi–layers of 3.25 and 3.75 nm are measured for [GaN/AlN]$_{20}$ (ESI Fig. S1) for [Al$_{0.35}$Ga$_{0.65}$N/Al$_{0.55}$Ga$_{0.45}$N]$_{20}$ (ESI Fig. S2) MQWs, respectively, in the supplementary information. The spectral lines centered around 570, 659 and 890 cm$^{-1}$ (Figs. 1(c) and 1(d)) are assigned to be the allowed symmetric Raman modes of GaN–$E_2^H$, AlN–$E_2^H$ and AlN–$A_1$(LO) modes of wurtzite phase, respectively.[22-35] The peak centered at ~750 cm$^{-1}$ is from the sapphire substrate. The mode is observed, as the MQW samples with wideband gap materials (band gap >3.4 eV at room temperature) are transparent to 514.5 nm (2.41 eV) excitation.

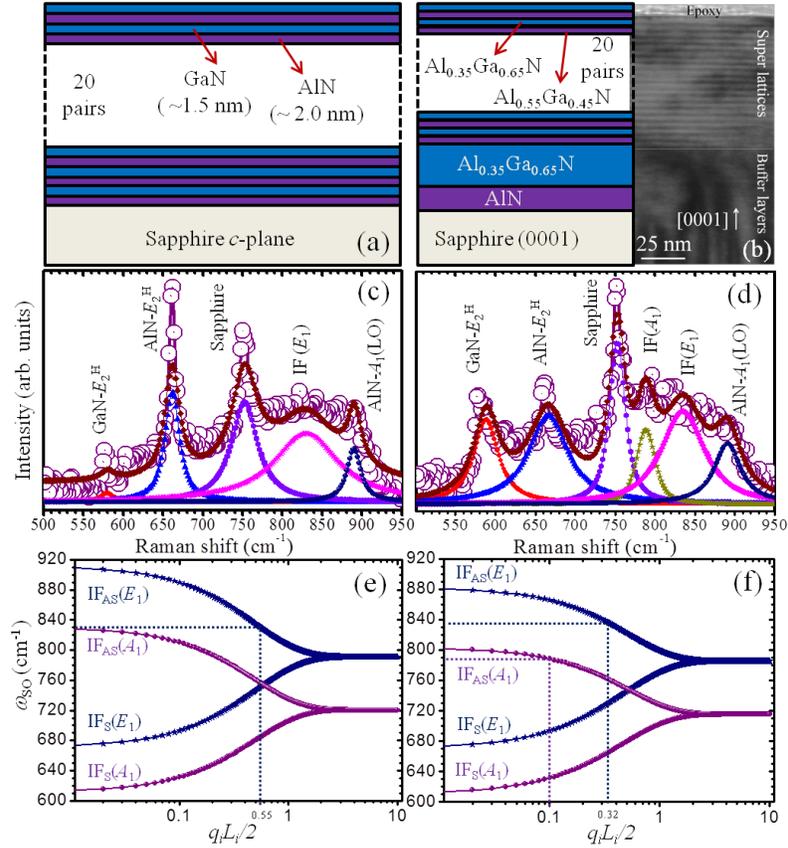

**Fig. 1**. The schematic representation of *c*–plane oriented MQW structures of (a) [GaN/AlN]$_{20}$ and (b) [Al$_{0.35}$Ga$_{0.65}$N/Al$_{0.55}$Ga$_{0.45}$N]$_{20}$ SLs. Outset shows a TEM micrograph of the MQWs. (c) and (d) the corresponding Raman spectra for 514.5 nm excitation recorded at 80K. (e) and (f) the respective simulated dispersion relation for the IF phonon modes. Dotted horizontal and vertical lines are drawn to indicate observed IF phonon modes and corresponding $q_iL_i/2$ values, respectively.

### 3.2 Computation of the dispersion relation for IF modes

For understanding the additional broad peaks (Figs 1(c) and 1(d)) observed around ~788 and ~830–834 cm$^{-1}$, we simulated the dispersion curves for the IF phonon modes by considering $\varepsilon_0$ and $\varepsilon_\infty$ as the respective static and high frequency value of dielectric constants ($\varepsilon(\omega)$) of the material. The symmetric (S) and asymmetric (AS) modes of IF phonon in 2D nanostructures are defined by the following equations,[15,23,26]

$$\omega_{IF}^2(q)_S = \omega_{TO}^2 \left[ \frac{\varepsilon_0 \tanh(q_i L_i/2) + \varepsilon_m}{\varepsilon_\infty \tanh(q_i L_i/2) + \varepsilon_m} \right] \quad (1)$$

$$\omega_{IF}^2(q)_{AS} = \omega_{TO}^2 \left[ \frac{\varepsilon_0 \coth(q_i L_i/2) + \varepsilon_m}{\varepsilon_\infty \coth(q_i L_i/2) + \varepsilon_m} \right] \quad (2)$$

Where $L_i$ ($i$ = 1, 2) is the edge width of the rectangular QW plates whose growth direction is considered to be along the z–direction; $q_i$ ($i$ = 1, 2) is the IF phonon wave–vectors in the QW structures; and $\varepsilon_m$ is the frequency dependent dielectric constant of the surrounding medium of the QW structures. In order to compute the dispersion relations (Eqns. (1) and (2)) for the IF phonon modes, it is assumed that the similar parity for the surface potentials of TO and LO phonon modes in the two mutually orthogonal directions ($q_1^2 + q_2^2 = q^2$; $q_1 L_1 = q_2 L_2$).[15,23,26]

The $\varepsilon_m$ follows the Lyddane–Sachs–Teller (LST) relation corresponding to $A_1$ and $E_1$ symmetries.[13,15] The values of $\varepsilon_0$ and $\varepsilon_\infty$ for AlN and GaN are considered from the available reports for assigning the dielectric contrast for the [GaN/AlN]$_{20}$ MQWs.[32] Similarly, the dielectric constants for the different 2D medium for [Al$_{0.35}$Ga$_{0.65}$N/Al$_{0.55}$Ga$_{0.45}$N]$_{20}$ structures are estimated on the basis of reported values of AlN and GaN,[32] followed by the numerical calculations with respect to the different Al percentage in the compositions of 2D layers in the MQW. The calculated values for $\varepsilon_0$ and $\varepsilon_\infty$ for each sample are shown in the Table 1. The corresponding frequency values of $A_1$(TO) and $E_1$(TO) symmetry modes for different MQW samples are 611 and 671 cm$^{-1}$, respectively, taken under consideration by the two–mode behavior of the random alloy model for AlGaN.[31-35]

**Table 1.** The dielectric constants for AlN, GaN and AlGaN with different stoichiometry.

| Materials | $\varepsilon_0$ | $\varepsilon_\infty$ |
|---|---|---|
| AlN | 8.50 | 4.60 |
| GaN | 8.90 | 5.40 |
| $Al_{0.55}Ga_{0.45}N$ | ~ 8.72 | ~ 5.04 |
| $Al_{0.50}Ga_{0.50}N$ | ~ 8.70 | ~ 5.00 |
| $Al_{0.35}Ga_{0.65}N$ | ~ 8.64 | ~ 4.88 |

With the help of the above tabulated data, the dispersion relations for IF phonon modes are simulated by using the Eqns. (1) as well as (2), and then the resultant curves are plotted for the respective sample as shown in the Figs. 1(e) and 1(f). The IF phonon mode for $[GaN/AlN]_{20}$ MQWs are calculated by considering AlN QWs as the material of interest and GaN QWs as the surrounding medium. Similarly, in the case of $[Al_{0.35}Ga_{0.65}N/Al_{0.55}Ga_{0.45}N]_{20}$ MQWs, $Al_{0.55}Ga_{0.45}N$ and $Al_{0.35}Ga_{0.65}N$ QWs are considered as the material of interest and the surrounding medium, respectively. The peak centered around ~830 $cm^{-1}$, observed in the Raman spectra of $[GaN/AlN]_{20}$ MQWs (Fig. 1(c)), may originate because of the presence of IF phonon mode of $E_1$ symmetry (IF($E_1$)) as indicated in the calculation. Similarly, the peaks centered at ~788 and ~834 $cm^{-1}$, observed in the Raman spectra of $[Al_{0.35}Ga_{0.65}N/Al_{0.55}Ga_{0.45}N]_{20}$ MQWs (Fig. 1(d)), may have evolved because of the simultaneous presence of IF phonon mode of $A_1$ symmetry (IF($A_1$)) as well as IF($E_1$) mode, respectively. The origin of the additional IF($A_1$) peak in the Raman spectra in the $[Al_{0.35}Ga_{0.65}N/Al_{0.55}Ga_{0.45}N]_{20}$ sample compared to that of the $[GaN/AlN]_{20}$ MQW may be because of the presence of additional $Al_{0.35}Ga_{0.65}N$(~320 nm)/AlN(~200 nm) buffer layer present on the sapphire substrate (Fig. 1(b) and ESI Fig. S2). The use of AlN buffer layer significantly reduces the dislocation density and improves the overall structural quality of the SL layers as well as the quantum efficiency of the MQW optoelectronic device.[8,29,30,36,37] In order to confirm it, a bi-layer thin film of $Al_{0.50}Ga_{0.50}N$(~300 nm)/AlN(~200 nm) (schematic Fig. 2(a)) was also studied for its symmetry allowed Raman phonon modes of AlN–$E_2^H$ and AlN–$A_1$(LO) along with the presence of IF modes (Fig. 2(b)). Similarly, the dispersion curve (Fig. 2(c)) for IF phonons are simulated by considering AlN layer as the material of interest and $Al_{0.50}Ga_{0.50}N$ as the surrounding medium using the corresponding values of $\varepsilon_0$ and $\varepsilon_\infty$ as shown in the Table 1. We could identify the mode frequencies of IF phonons, at IF($A_1$) ~794 $cm^{-1}$ with significant intensity and IF($E_1$) at ~846 $cm^{-1}$ (Fig. 2(b)). The distinct IF($A_1$) mode with high intensity, observed for $Al_{50}Ga_{50}N$/AlN bi-layer sample, is close to the observed value of ~788 $cm^{-1}$ in case of $[Al_{0.35}Ga_{0.65}N/Al_{0.55}Ga_{0.45}N]_{20}$ MQW sample. The small shift of ~6 cm$^{-1}$ for IF($A_1$) and ~13 cm$^{-1}$ for IF($E_1$) modes, observed in the Al$_{50}$Ga$_{50}$N/AlN bi–layer thin film as compared to those for [Al$_{0.35}$Ga$_{0.65}$N/Al$_{0.55}$Ga$_{0.45}$N]$_{20}$ MQW, may be because of the different dielectric constants in the layered structures with different Al contents (Table 1). Moreover, the broadening of IF($E_1$) mode corresponding to the [Al$_{0.35}$Ga$_{0.65}$N/Al$_{0.55}$Ga$_{0.45}$N]$_{20}$ sample (Fig. 1(d)) may be because of the contribution originating from the IF($E_1$) mode corresponding to the Al$_{0.35}$Ga$_{0.65}$N/AlN buffer layer. In this regard presence of a broad peak at ~846 cm$^{-1}$ corresponding to the Al$_{50}$Ga$_{50}$N/AlN bi–layer sample may be noted (Fig. 2(b)).

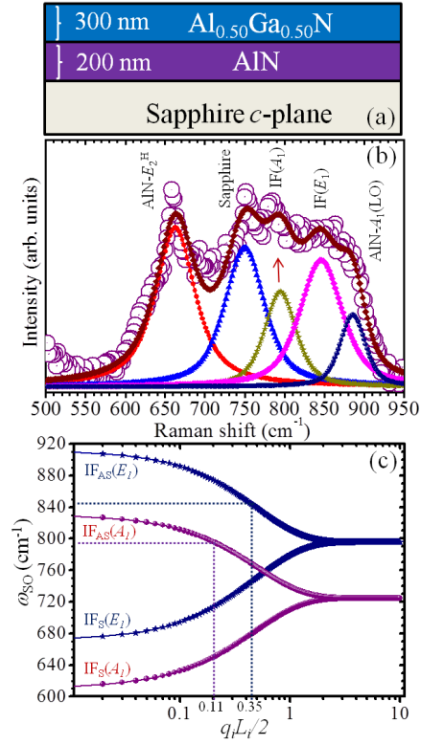

**Fig. 2.** (a) The schematic representation of thin film structures of Al$_{0.50}$Ga$_{0.50}$N/AlN (b) The Raman spectra for 514.4 nm excitation recorded at 80K and (c) the respective simulated dispersion relation for the IF phonon modes. Dotted horizontal and vertical lines are drawn to indicate observed IF phonon modes and corresponding $q_iL_i/2$ values, respectively.

### 3.3 Origin for the IF modes probed by Raman spectra

In polar crystals like group III nitride, the SO modes are observed on the lower wave number side of the (LO)$_{q=0}$ mode and higher wave number side of the (TO)$_{q=0}$ mode.[12-22] The periodic surface modulation and interfaces of the SL structures of 2D MQWs (ESI Figs. S1 and S2) may lead to the relaxation of Raman selection rules because of the breakdown of translational symmetry of the surface potential,[12,26] and it is anticipated that the unidentified Raman modes observed in the recorded spectra from the MQWs belong to the above mentioned IF($A_1$) and IF($E_1$) phonon modes, respectively. The surface layers of MQWs and buffer layers absorbs an additional momentum in terms of the quantized units of $q$ (= $2\pi/\lambda$) and the expected edge lengths ($L$) corresponding to the IF($A_1$) and IF($E_1$) phonon modes are calculated for a periodicity of the surrounding medium separated with a thickness of the material of interest, $\lambda$ ~2 nm (for QWs) and $\lambda$ ~200 nm (for buffer layer and bi–layer thin film) (Table 2).

The nature of IF($A_1$) mode, observed in the [$Al_{0.35}Ga_{0.65}N/Al_{0.55}Ga_{0.45}N$]$_{20}$ SL, is attributed to the presence of MQW structure as well as the $Al_{0.35}Ga_{0.65}N$/AlN buffer layers present in it. Therefore, in case of the sample [$Al_{0.35}Ga_{0.65}N/Al_{0.55}Ga_{0.45}N$]$_{20}$, the corresponding $q$ and $L$ values are calculated for IF($A_1$) modes for a pair of layer thickness, $\lambda$ ~2 and ~200 nm, respectively. The observed IF($A_1$) mode wave number with significant intensity as well as the order of $q$ and $L$ (Table 2) match well with the sample which posses $Al_{0.50}Ga_{0.50}N$/AlN bi–layered thin film alone on the sapphire substrate (Fig. 2). The 2D shapes with finite $L$ can be formed on the sample surface by the horizontal uneven irregularities generated during the growth period. The typical FESEM image of MQWs shows (ESI Fig. S3) the nanometer sized platelets uniformly distributed all over the horizontal surface of SLs and the surface protrusions are of the order of 10s to 100s of nm. The integral multiples of nanometer sized protrusions also can act as a micron sized platelets (~2 μm). Moreover, the required $L$ values fall in the order of 10s and 100s of nanometers to a few micrometers (~2 μm). It is very important to notice that the spot size of the excitation laser (~2 μm) can easily cover the required area generated by the 2D plate lengths.

**Table 2.** The $q$ and $\lambda$ of the respective IF phonon modes observed in the Raman spectra of the MQW structures with an expected 2D plate length ($L$).

| Substrate for all of the sample is sapphire | [GaN(~1.25 nm)/AlN(~2.0 nm)]$_{20}$ | | [Al$_{0.35}$Ga$_{0.65}$N(~1.75 nm)/Al$_{0.55}$Ga$_{0.45}$N(~2.0 nm)]$_{20}$ on Al$_{0.35}$Ga$_{0.65}$N(~300 nm)/ AlN(~200 nm) buffer layer | | Al$_{0.50}$Ga$_{0.50}$N(~320 nm)/AlN(~200 nm) | |
|---|---|---|---|---|---|---|
| IF modes | $\omega_{IF}(A_1)$ | $\omega_{IF}(E_1)$ | $\omega_{IF}(A_1)$ | $\omega_{IF}(E_1)$ | $\omega_{IF}(A_1)$ | $\omega_{IF}(E_1)$ |
| Wave number | - | 830 | 788 cm$^{-1}$ | 834 cm$^{-1}$ | 794 cm$^{-1}$ | 846 cm$^{-1}$ |
| $x = qL/2$ | - | 0.55 | 0.10 | 0.32 | 0.11 | 0.35 |
| $\lambda$ (periodicity of surrounding medium) | - | 2 nm | 200 nm (buffer layer) | 2 nm (QWs) 200 nm (buffer layer) | 200 nm (thin film) | 2 nm (QWs) 200 nm (thin film) |
| $q = 2\pi/\lambda$ | - | 3.14 × 10$^7$ cm$^{-1}$ | 3.14 × 10$^5$ cm$^{-1}$ | 3.14 × 10$^7$ cm$^{-1}$ and 3.14 × 10$^5$ cm$^{-1}$ | 3.14 × 10$^5$ cm$^{-1}$ | 3.14 × 10$^7$ cm$^{-1}$ |
| $L = 2x/q$ (nm) | - | 35 | 637 | 20.38 ($\lambda$ ~2 nm) 2038 ($\lambda$ ~200 nm) | 700 | 22.40 ($\lambda$ ~2 nm) 2240 ($\lambda$ ~200 nm) |

The small shift in IF($E_1$) phonon mode (~4 cm$^{-1}$) for [Al$_{0.35}$Ga$_{0.65}$N/Al$_{0.55}$Ga$_{0.45}$N]$_{20}$ compared to that for the [GaN/AlN]$_{20}$ MQWs is because of the difference in $\varepsilon_m$ of the surrounding medium. This red shift of the observed mode in between the allowed (TO)$_{q=0}$ and (LO)$_{q=0}$ Raman modes with increasing $\varepsilon_m$ value because of the change in the surrounding medium can be considered as a further confirmation for the presence of IF mode.[12-22,26] We also studied the temperature dependent evolution of IF phonon modes along with the allowed Raman modes of MQW structures (ESI Fig. S4). From the temperature dependent Raman studies of MQW structures, it is observed that the relative strength of IF modes along with allowed Raman modes are also increased with decrease in temperature.

## 4. Conclusions

In summary, apart from the symmetry allowed Raman modes, the interface (IF) phonon modes are also observed in the Raman spectra of [GaN(~1.25 nm)/AlN (~2.0 nm)]$_{20}$ and [Al$_{0.35}$Ga$_{0.65}$N(~1.75 nm)//Al$_{0.55}$Ga$_{0.45}$N(~2.0 nm)]]$_{20}$ multi quantum well (MQW) structures. The dispersion relation for both IF($A_1$) and IF($E_1$) phonon modes for MQW structures are calculated and are matched with the experimentally observed IF phonon modes in the Raman spectra. For [GaN/AlN]$_{20}$ MQW structure, the IF($E_1$) phonon mode alone is observed, whereas in case of [Al$_{0.35}$Ga$_{0.65}$N/Al$_{0.55}$Ga$_{0.45}$N]$_{20}$ MQW structure, both IF($A_1$) and IF($E_1$) are identified simultaneously. The red shift of IF mode (~4 cm$^{-1}$) of [GaN/AlN]$_{20}$ compared to that for [Al$_{0.35}$Ga$_{0.65}$N/Al$_{0.55}$Ga$_{0.45}$N]$_{20}$ MQW structure with the change in dielectric constant of the surrounding medium, further confirms the existence of IF modes in the observed Raman Spectra. The relative increase for intensity of IF($A_1$) in [Al$_{0.35}$Ga$_{0.65}$N/Al$_{0.55}$Ga$_{0.45}$N]$_{20}$ MQW sample is evolved because of the presence of Al$_{0.35}$Ga$_{0.65}$N/AlN buffer layer in the sample, as confirmed with the observation of the similar IF($A_1$) mode with higher intensity in the Al$_{0.50}$Ga$_{0.50}$N/AlN bi–layer thin film structure. The observation of the IF phonon modes is attributed to the relaxation of Raman selection rules because of the breakdown of translational symmetry because of the presence of the periodic interfaces and surface modulations of SL structure of MQWs. The integral multiples of the nanometer sized protrusions, uniformly distributed all over the horizontal surface of SLs, in the order of 10s and 100s of nanometers to a few micrometers (~2 μm) are found to act as the expected edge lengths responsible for the breakdown of symmetry.

## Acknowledgements

One of us (AKS) acknowledges the Department of Atomic Energy, India for allowing him to continue the research work. We thank N. Meher of Materials Physics Division, S. Parida and R. Basu of SND, IGCAR for their valuable suggestions and help in theoretical simulations. One of us (CS) also likes to acknowledge the Department of Science and Technology (DST) INSPIRE Fellowship for funding his research. The work at the University of Calcutta work was partially funded by the Department of Information Technology (12(3)/2011-PDD), Government of India and the MHRD, Government of India.

**Notes and references**

Electronic Supplementary Information (ESI);

# Interface Phonon Modes in the [AlN/GaN]$_{20}$ and [Al$_{0.35}$Ga$_{0.65}$N/Al$_{0.55}$Ga$_{0.45}$N]$_{20}$ 2D Multi Quantum Well Structures

A. K. Sivadasan,[*,1] Chirantan Singha,[2] A. Bhattacharyya,[3] and Sandip Dhara[*,1]

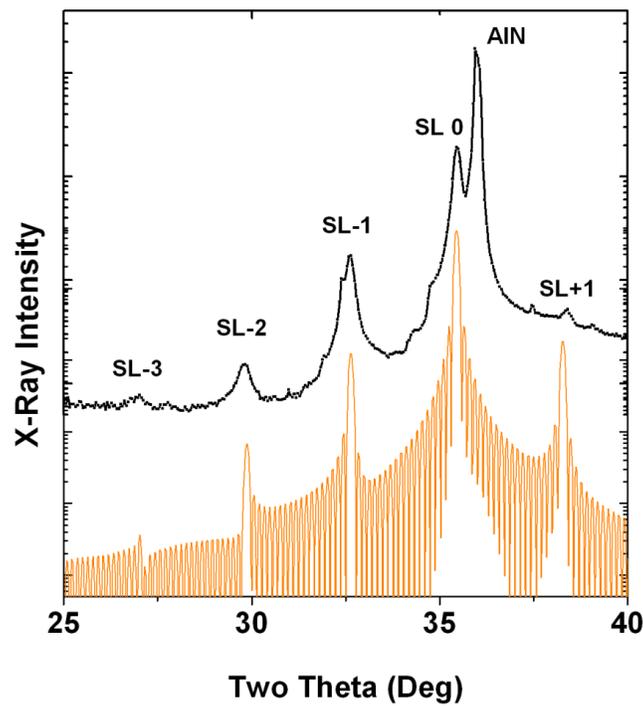

**Fig. S1**. High resolution x-ray diffraction data of [AlN/GaN]$_{20}$ MQWs showing the AlN layer and several superlattice peaks. Simulation using the X-ray Kinematic model [1] shows that the periodicity of the repeating bi–layer of QW structures is about 3.25 nm.

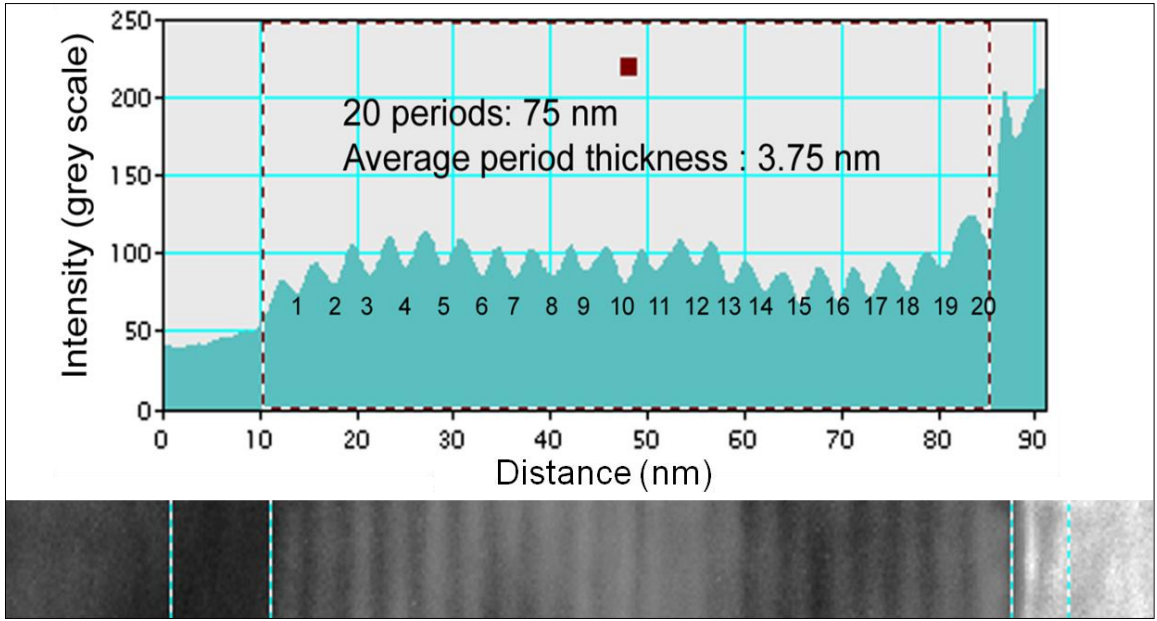

**Fig. S2.** Transmission electron microscopic image and corresponding intensity plot of $[Al_{0.35}Ga_{0.65}N/Al_{0.55}Ga_{0.45}N]_{20}$ MQW showing 20 periods of repeating bi–layer of QWs with 3.75 nm.

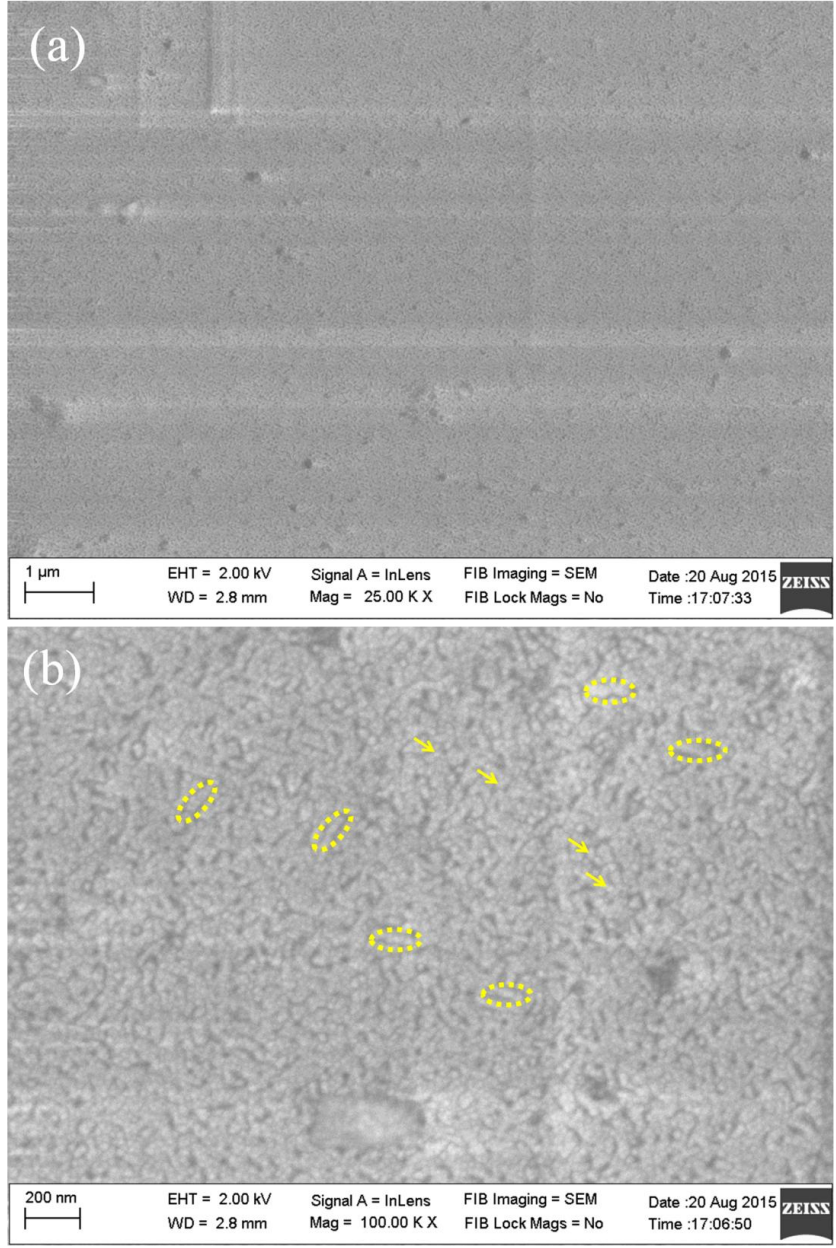

**Fig. S3.** (a) Typical FESEM image showing morphology of MQWs containing the nanometer sized platelets uniformly distributed all over the surface (~10×10 µm). (b) The high magnification image for the surface protrusions in the order of 10s (arrow head) to 100s (encircled portions) of nm. The integral multiples of nanometer sized protrusions also can act as a micron sized platelets ( ~2 µm).

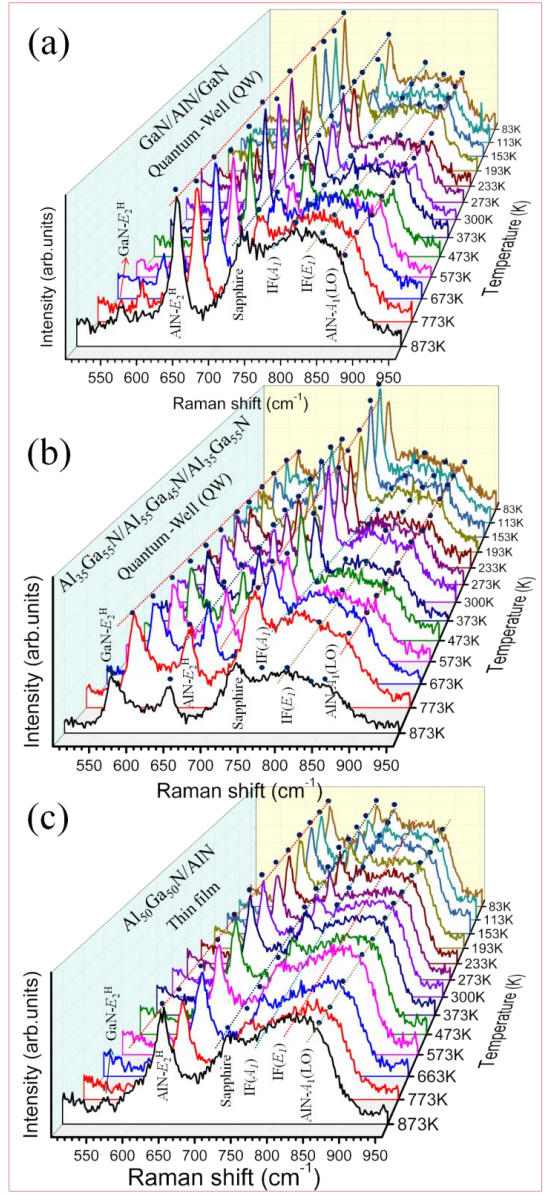

**Fig. S4** The temperature dependent evolution of IF phonon modes along with allowed Raman modes of (a) [GaN/AlN]$_{20}$ MQWs (b) [Al$_{0.35}$Ga$_{0.65}$N/Al$_{0.55}$Ga$_{0.45}$N]$_{20}$ MQWs and (c) Al$_{0.50}$Ga$_{0.50}$N/AlN thin film.